\documentclass[a4paper]{article} 
\usepackage{epsfig,amssymb}
\usepackage{amsmath}
\usepackage{natbib}
\usepackage{graphics}% Include figure files
\usepackage{graphicx}% Include figure files
\usepackage{dcolumn}% Align table columns on decimal point
\usepackage{bm}% bold math
\usepackage{epsfig}

\begin{document}
%******************************************************************
%\baselineskip=.33in
%******************************************************************

\begin{center}

\Large{\bf Lunar Laser Ranging Contributions
to Relativity and Geodesy}

\vspace{15pt}

\normalsize
\bigskip 

J\"urgen M\"uller$^a$, James G.\ Williams$^b$, and Slava G.\ Turyshev$^b$

\normalsize
\vskip 10pt

{\it 
$^a$Institut f\"{u}r Erdmessung (IfE), University of Hannover, Schneiderberg 50, 30167 Hannover, Germany\\
mueller@ife.uni-hannover.de\\
$^b$Jet Propulsion Laboratory, California Institute of Technology, \\
4800 Oak Grove Drive, Pasadena, CA 91109, USA
}
\vspace{0.25in}
\end{center}

%**************************************************
\begin{abstract}
Lunar laser ranging (LLR) is used to conduct high-precision measurements of ranges between an observatory on Earth and a laser retro-reflector on the lunar surface. Over the years, LLR has benefited from a number of improvements both in observing technology and data modeling, which led to the current accuracy of post-fit residuals of $\sim 2$~cm. Today LLR is a primary technique to study the dynamics of the Earth-Moon system  and is especially important for gravitational physics, geodesy and studies of the lunar interior. 
When the gravitational physics is concerned, LLR is used to perform high-accuracy tests of the equivalence principle, to search for a time-variation in the gravitational constant, and to test predictions of various alternative theories of gravity. The gravitational physics parameters cause both secular and periodic effects on the lunar orbit that are detectable with the present day LLR; in addition, the accuracy of their determination benefits from the 35 years of the LLR data span.  
On the geodesy front, LLR contributes to the determination of Earth orientation parameters, such as nutation, precession (including relativistic precession), polar motion, and UT1, i.e.\ especially to the long-term variation of these effects. LLR contributes to the realization of both the terrestrial and selenocentric reference frames.  The realization of a dynamically defined inertial reference frame, in contrast to the kinematically realized frame of VLBI, offers new possibilities for mutual cross-checking and confirmation. Finally, LLR also investigates the processes related to the Moon's interior dynamics. 
Here, we review the LLR technique focusing on its impact on relativity and give an outlook to further applications, e.g. in geodesy. We present results of our dedicated studies to investigate the sensitivity of LLR data with respect to the relativistic quantities; we also present the computed corresponding spectra indicating the typical periods related to the relativistic effects. We discuss the current observational situation and the level of LLR modeling implemented to date. We emphasis the need for the modeling improvement for the near future LLR opportunities. We also address improvements needed to fully utilize the scientific potential of LLR.

\vspace{0.3cm}\noindent

\end{abstract}

%**************************************************
\section{Introduction}
Being one of the first space geodetic techniques, lunar laser ranging (LLR) has routinely provided observations for more than 35 years.  The LLR data are collected as normal points, i.e.\ the combination of lunar returns obtained over a short time span of 10 to 20 minutes. Out of $\approx 10^{19}$ photons sent per pulse by the transmitter, less than 1 is statistically detected at the receiver \cite{wil96}; this is because of the combination of several factors, namely energy loss (i.e. $1/R^4$ law), atmospherical extinction and geometric reasons (rather small telescope apertures and reflector areas). Moreover, the detection of real lunar returns is rather difficult as dedicated data filtering (spatially, temporally and spectrally) is required. These conditions are the main reason, why only a few observatories worldwide are capable of laser ranging to the Moon.
  
Observations began shortly after the first Apollo 11 manned mission to the Moon in 1969 which deployed a passive retro-reflector on its surface. Two American and two French-built reflector arrays (transported by Soviet spacecraft) followed until 1973.\footnote{One of the reflector-arrays (of the Soviet Luna 17 mission, see also Fig.\ \ref{refmoon}) has not been tracked operationally. The reason could be that the coordinates are not known well enough or that the reflectors are covered by dust or the transport cap.} Most observations are taken to the largest reflector array, that of the Apollo 15 mission. Over the years more than 16,000 LLR measurements have by now been made of the distance between Earth observatories and lunar reflectors. Most LLR data have been collected by the Observatoire de la C\^ote d'Azur (OCA, France), the McDonald observatory (Texas, USA) and - until 1990 - Haleakala (Hawaii, USA). The new data are still coming, but today only the first two stations operate regularly. Understanding unexpected and small effects is very difficult with only one or two operating stations, because possible instrumental systematics of the ranging system can not be separated from real scientific effects reliably. In order to further increase the impact of LLR in Relativity and Earth sciences more stations, with a wide geographical distribution, are needed. Therefore the Italian colleagues have set up a new site in Matera which has provided first LLR data quite recently. A new site with lunar capability is currently being built at the Apache Point Observatory, New Mexico, USA.  This station, called APOLLO, is designed for a mm-level accuracy ranging \citep{Murphy_etal_2000,wil04b}. However, to fully exploit the available LLR potential, a few more sites capable of tracking the Moon are needed, especially at diverse locations including the Southern hemisphere.

%*********************************
\begin{figure*}[t!]
\begin{minipage}[b]{.46\linewidth}
\centering \psfig{width=7.4cm,     file=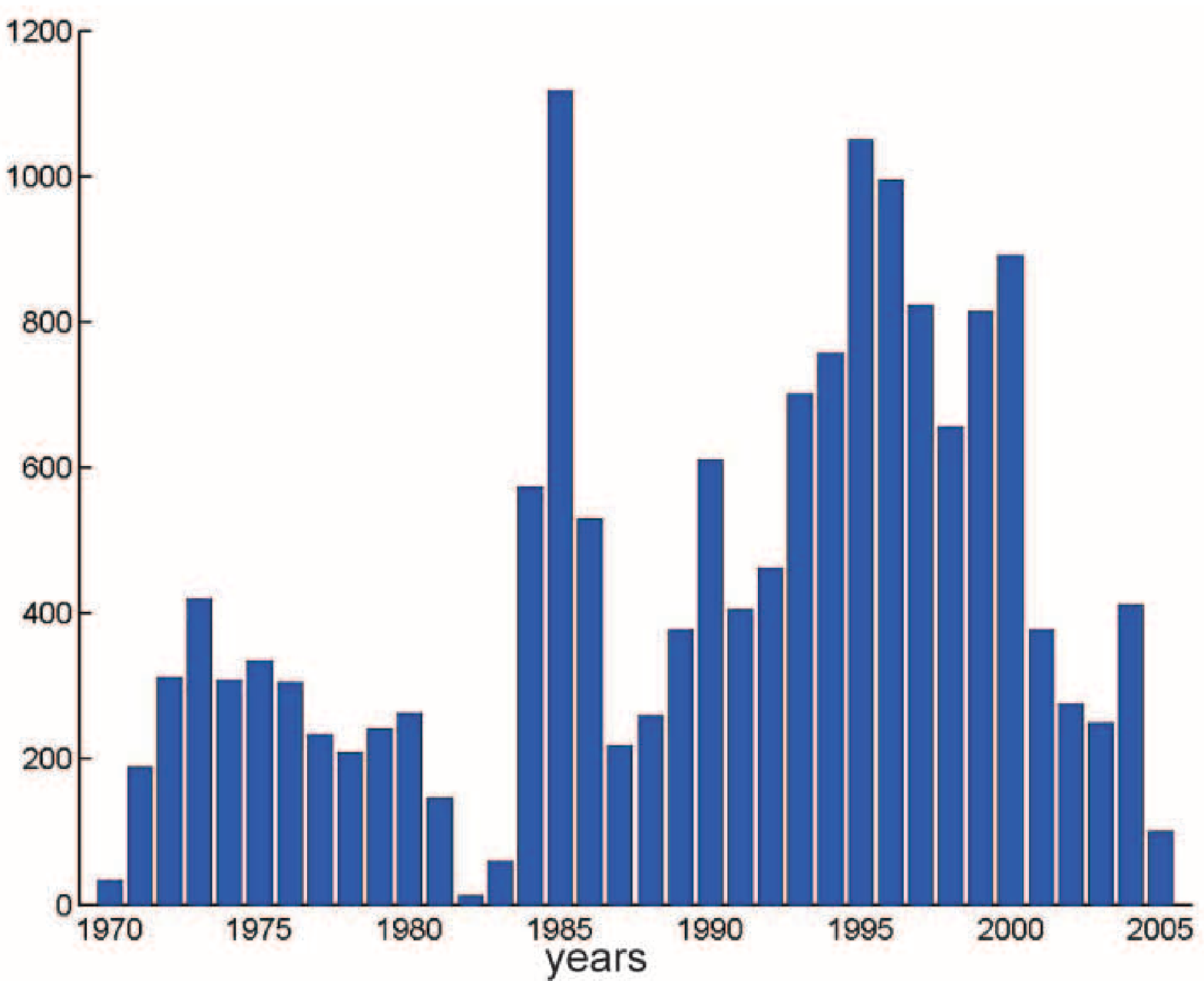}
\end{minipage} 
\hfill
\begin{minipage}[b]{.46\linewidth}
\centering \psfig{width=7.4cm,     file=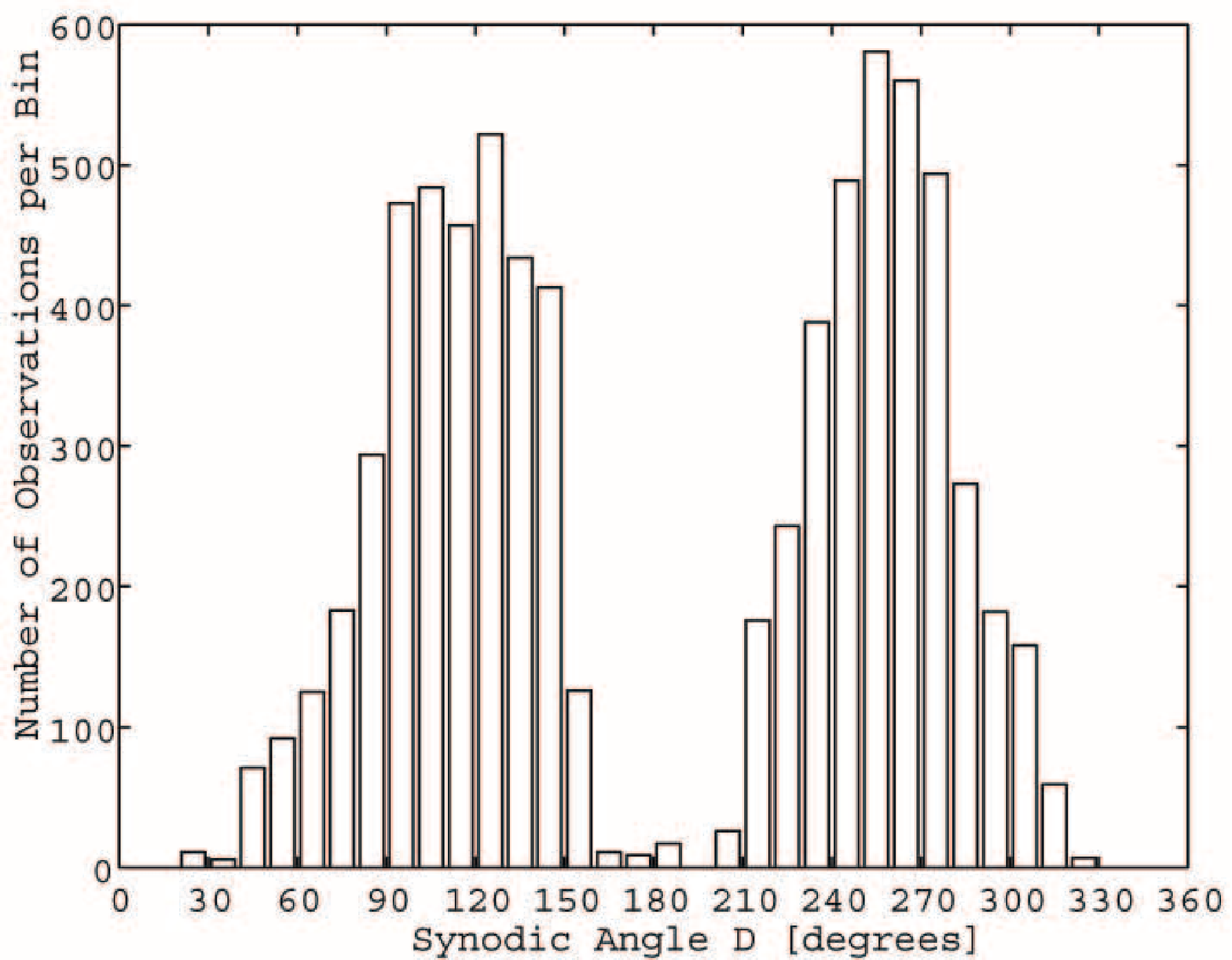}
\end{minipage}
\caption{(a) Left: Lunar observations per year, 1970 - 2005. (b) Data distribution as a function of the synodic angle $D$.
 \label{ann_syn}}
\vskip -5pt 
\end{figure*}
%*********************************

\begin{figure*}[ht]
\begin{minipage}[b]{.46\linewidth}
\centering \psfig{width=7.4cm,     file=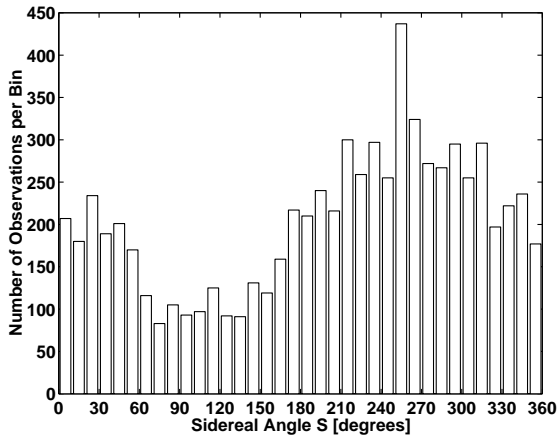}
\end{minipage} 
\hfill
\begin{minipage}[b]{.46\linewidth}
\centering \psfig{width=7.4cm,     file=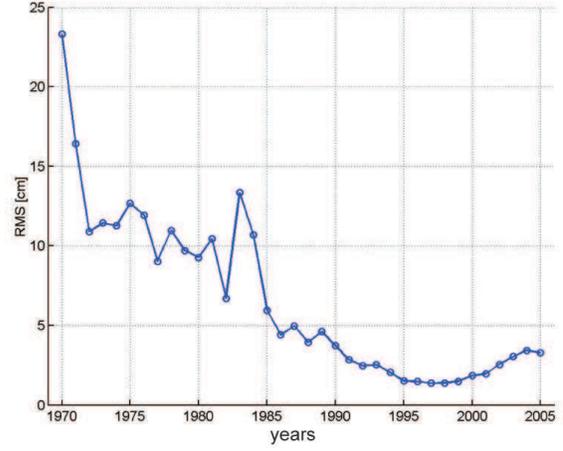}
\end{minipage}
\caption{(a) Data distribution as a function of the  sidereal angle $S,$ where the Moon is south of equator from 0$^\circ$ to 180$^\circ$. 
(b) Weighted residuals (observed-computed Earth-Moon distance) annually averaged. \label{sid_rms}}
\vskip -5pt 
\end{figure*}

Fig.~\ref{ann_syn}a shows the number of LLR normal points per 
year since 1970. As shown there and in Fig.~\ref{ann_syn}b, the range data have not been accumulated uniformly; substantial variations in data density exist as a function of synodic angle~$D$, these phase angles are represented by 36 bins of 10 degree width. In  Fig.~\ref{ann_syn}b, data gaps are seen near new Moon (0 and 360 degrees) and full Moon (180 degrees) phases, and asymmetry about quarter Moon (90 and 270 degrees) phases also is exhibited.  The former properties of this data distribution are a consequence of operational restrictions, such as difficulties to  operate near the bright sun in daylight (i.e.\ new Moon) or of high background solar illumination noise (i.e.\ full Moon). Note also asymmetry about quarter  Moon phases. The uneven distribution with respect to the lunar sidereal angle shown in Fig.~\ref{sid_rms}a represents the increased difficulty of making observations from northern hemisphere observatories to the Moon when it is located over the southern hemisphere. Here, the situation will only be improve if a southern observatory (e.g.\ in Australia) will start to track the Moon. 
It might be possible that new missions to the Moon could be helpful in this respect; the deployment of active laser transponders could allow satellite laser ranging systems to participate in LLR.

%*********************************
%*********************************

While measurement precision for all model parameters benefit from the ever-increasing improvement in precision of individual range measurements (which now is at the few cm level, see also  Fig.\ \ref{sid_rms}b), some parameters of scientific interest, such as time variation of Newton's coupling parameter $\dot G/G$ or precession rate of lunar 
perigee, particularly benefit from the long time period (35 years and growing) of range measurements.

In the 1970s LLR was an early space technique for determining Earth orientation parameters (EOP). Today LLR still competes with other space geodetic techniques, and because of large improvements in ranging precision (30 cm in 1969 to $\sim2$~cm today), it now serves as one of the strongest tools in the solar system for testing general relativity. Moreover, parameters such as the station coordinates and velocities 
contributed to the International Terrestrial Reference Frame ITRF2000.  EOP quantities were used in combined solutions of the International Earth Rotation and Reference Systems Service IERS ($\sigma$~=~0.5~mas).

\section{LLR Model and Relativity}

The existing LLR model at IfE has been developed to compute the LLR observables --- the roundtrip travel times of laser pulses between stations on the Earth and passive reflectors on the Moon (see e.g.\ M\"uller et al.\ 1996, M\"uller and Nordtvedt 1998 or M\"uller 2000, 2001, M\"uller and Tesmer 2002, Williams et al.\ 2005b and the references therein). The model is fully relativistic and is complete up to first post-Newtonian ($1/c^2$) level; it uses the Einstein's general theory of relativity -- the standard theory of gravity. The modeling of the relativistic parts is much more challenging than, e.g., in SLR, because the relativistic corrections increase the farther the distance becomes. The basic observation equation reads
\begin{equation}
d = c \frac{\tau}{2} = \left| \bf r_{\rm em} - r_{\rm station} + r_{\rm reflector}\right|+ c\; \Delta\tau
\label{eq:1}
\end{equation}
where $d$ is the station-reflector distance, $c$ the speed of light, $\tau$ the pulse travel time, $\bf r_{\rm em}$ the vector connecting the geocenter and the selenocenter, $\bf r_{\rm station}$ the geocentric position vector of the observatory, and $\bf r_{\rm reflector}$ the selenocentric position vector of the reflector arrays. $\Delta\tau$ describes corrections of the travel time caused by atmospheric effects, but also (relativistic) transformations into the right time system as well as the light time equation (Shapiro effect). In order to apply Eq.~(\ref{eq:1}), all vectors have to be transformed in one common reference frame (in our case the inertial frame) which requires consistent relativistic transformations, so-called pseudo-Lorentz transformations. The Earth-Moon vector $\bf r_{\rm em}$ can only be obtained by numerical integration of the corresponding equation of motion (here in simplified version):
\begin{eqnarray}
{\bf \ddot r_{\rm em}} =  - \frac{GM_{\rm e+m}}{r_{\rm em}^3}{\bf r_{\rm em}}  + GM_s \left( \frac{\bf r_{\rm s} - r_{\rm em}}{\left| {\bf r_{\rm s} - r_{\rm em}}\right|^3} - \frac{\bf r_{\rm s}}{r_{\rm s}^3} \right) 
  + {\bf b_{\rm newtonian}} + {\bf b_{\rm relativistic}} .
\end{eqnarray}
$\bf \ddot r_{\rm em}$ is the relative acceleration vector between Earth and Moon, $GM_{\rm e+m}$ is the Earth-Moon mass times the gravitational constant, $\bf r_{\rm s}$ the geocentric position vector of the Sun, $r_{\rm s}$ the Earth-Sun distance, $M_s$ the solar mass, ${\bf b_{\rm newtonian}}$ comprises all further Newtonian terms like the effect of the other planets or the gravitational fields of Earth and Moon, and $\bf b_{\rm relativistic}$ indicates all 'relativistic' terms, i.e.\ those entering the Einstein-Infeld-Hofmann (EIH) equations. Corresponding relativistic equations are applied to describe the rotational motion of the Moon. The rotation angles are then applied to rotate the selenocentric reflector coordinates of Eq.\ (1) into the inertial frame. For the transformation of the geocentric station coordinates, the well known rotation matrices (precession, nutation, GAST, etc.) are used, see IERS 2003.

The modeling of the 'Newtonian' parts has been set up according to IERS Conventions (IERS 2003), but it is restricted to the 1 cm level.  The weights are based upon the accuracy estimates for the normal points as provided by the observatories. Based upon this model, two groups of parameters (170 in total) are determined by a weighted least-squares fit of the observations. The first group includes the so-called  'Newtonian' parameters such as
\begin{itemize}
\item geocentric coordinates of three Earth-based LLR stations and their velocities;
           
\item a set of EOPs (luni-solar precession constant, nutation coefficients of the 18.6 years period, Earth's rotation 
UT0 and variation of latitude by polar motion);

\item selenocentric coordinates of four retro-reflectors;

\item rotation of the Moon at one initial epoch (physical librations);

\item orbit (position and velocity) of the Moon at this epoch;

\item orbit of the Earth-Moon system about the Sun at one epoch;

\item mass of the Earth-Moon system times the gravitational constant;

\item the lowest mass multipole moments of the Moon;

\item lunar Love number and a rotational energy dissipation parameter;

\item lag angle indicating the lunar tidal acceleration responsible for the increase of the Earth-Moon distance
(about 3.8 cm/yr), the increase in the lunar orbit period and the slowdown of Earth's angular velocity.
\end{itemize}

The second group of parameters is used to perform LLR tests of viable modifications of the general theory of relativity. The general theory of relativity is not expected to be perfect, because Einstein's theory and quantum mechanics are fundamentally incompatible. Therefore, it is important in physics to find out at what level of accuracy it fails. Relativistic  parameters to be determined by LLR analyses are (values for general relativity are given in parentheses):

\begin{enumerate}
\item Strong equivalence principle (EP) parameter $\eta$, which for metric theories is $\eta=4\beta-3-\gamma$ (= 0).
In LLR a violation of the equivalence principle would show up as a displacement of the lunar orbit along the direction to the Sun. LLR is the dominant test of the strong equivalence principle, i.e.\ for self-gravitating bodies. 

\item Space-curvature parameter $\gamma$ (= 1) and non-linearity parameter $\beta$ (= 1). LLR also has the capability of determining the PPN parameters $\beta$ and $\gamma$ directly from the point-mass 
orbit perturbations (i.e.\ as described by the EIH equations), but, e.g., $\beta$ may be derived much better by combining the EP parameter $\eta$ with an independent determination of $\gamma$ (see below).

\item Time variation of the gravitational coupling parameter $\dot G/G$ (= 0  $\rm yr^{-1}$). This is the second most important gravitational physics result that LLR provides. Einstein's theory does not predict a changing $G$, but some other theories do. So it is important to measure this as well as possible. The sensitivity improves like the square of the data span.  
 
\item Geodetic de {Sitter} precession ${\Omega}_{\rm dS}$  of 
the lunar orbit ($\simeq 1.92$~"/cy). LLR has also provided the only accurate determination of the geodetic precession. The dedicated space mission GP-B will provide an improved accuracy, if that mission is successfully completed. 

\item Coupling constant $\alpha$~(=~0) of Yukawa potential for the Earth-Moon distance which corresponds to a test of Newton's inverse square law. 

\item $\alpha_1$ (= 0) and $\alpha_2$ (= 0) which parametrize 'preferred frame' effects in metric gravity.

\item Combination of parameters $\zeta_1 - \zeta_0 - 1$ (= 0) derived in the Mansouri and Sexl (1977) formalism indicating a violation of special relativity (there: Lorentz contraction parameter $\zeta_1 = 1/2$, time dilation parameter $\zeta_0 = - 1/2$).

\item EP-violating coupling of normal matter to 'dark matter' at the galactic center. It is a similar test to item 1 above, but now different periods are affected (mainly the sidereal month).

\item A further application is the detection of the Sun's $J_2$ ($\approx 10^{-7}$) from LLR data (independent to other methods as solar seismology), which affects the anomalous perihelion shift of Mercury, one of the classical tests of relativity. The present uncertainty ($\approx 10^{-6}$) is larger than the expected value. The parameter  $J_2$ is not further discussed in this paper and will be addressed in more detail in an upcoming study.
\end{enumerate}

The determination of the relativistic parameters  indicated above can  be accomplished either by modifying the equations of motion (i.e.\ parametrizing present terms or adding new ones) or by deriving analytical expressions for their effect on the Earth-Moon distance. In the first case the needed partial derivatives can be computed by
numerical differentiation, in the second case by direct derivation of the analytical terms.

Many relativistic effects produce a sequence of periodic perturbations of the Earth-Moon range
\begin{equation}
\Delta r_{EM} = \sum_{i=1}^n \Big[A_i \cos (\omega_i \Delta t + \Phi_i)+B_i \Delta t\sin (\omega_i \Delta t + \Phi_i)\Big]. 
\end{equation}
$A_i$ and $B_i,\; \omega_i,$ and $\phi_i$ are the amplitudes, frequencies, and phases, respectively, of the various perturbations. Some sample periods of perturbations important for the measurement of various parameters are given in Table~1.\footnote{Note: the designations should not be used as formulae for the computation of the corresponding periods, e.g.\ the period `sidereal-2$\cdot$annual' has to be calculated as $1/(1/27.32^d - 2/365.25^d) \approx 32.13^d$. In addition,  `secular + emerging periodic' means the changing orbital frequencies induced by $\dot G/G$ are starting to become better signals than the secular rate of change of the Earth-Moon range in LLR.}
{}
\begin{table}[bht]
\caption{Typical periods of some relativistic quantities, taken from M\"uller et al.\ (1999).}
\begin{center}
\begin{tabular}{|c|c|}
\hline
Parameter & Typical Periods \\ \hline\hline  
$\eta$ &  synodic (29$^d$12$^h$44$^m$2.9$^s$)\\ \hline
$\dot G/G$ & secular + emerging periodic  \\ \hline
$\alpha_1$ & sidereal, annual, sidereal-2$\cdot$annual,\\
           & anomalistic\ (27$^d$13$^h$18$^m$33.2$^s$) $\pm$ annual, synodic\\ \hline
$\alpha_2$ & 2$\cdot$sidereal, 2$\cdot$sidereal-anomalistic, nodal (6798$^d$) \\ \hline
$\zeta_1-\zeta_0-1$ & annual (365.25$^d$)\\ \hline
$\delta g_{\rm galactic}$ & sidereal (27$^d$7$^h$43$^m$11.5$^s$)\\ \hline
\end{tabular}
\end{center}
\vskip -10pt
\end{table}

%*********************************
\begin{figure*}[ht]
\begin{minipage}[b]{.46\linewidth}
\centering \psfig{width=7.4cm,     file=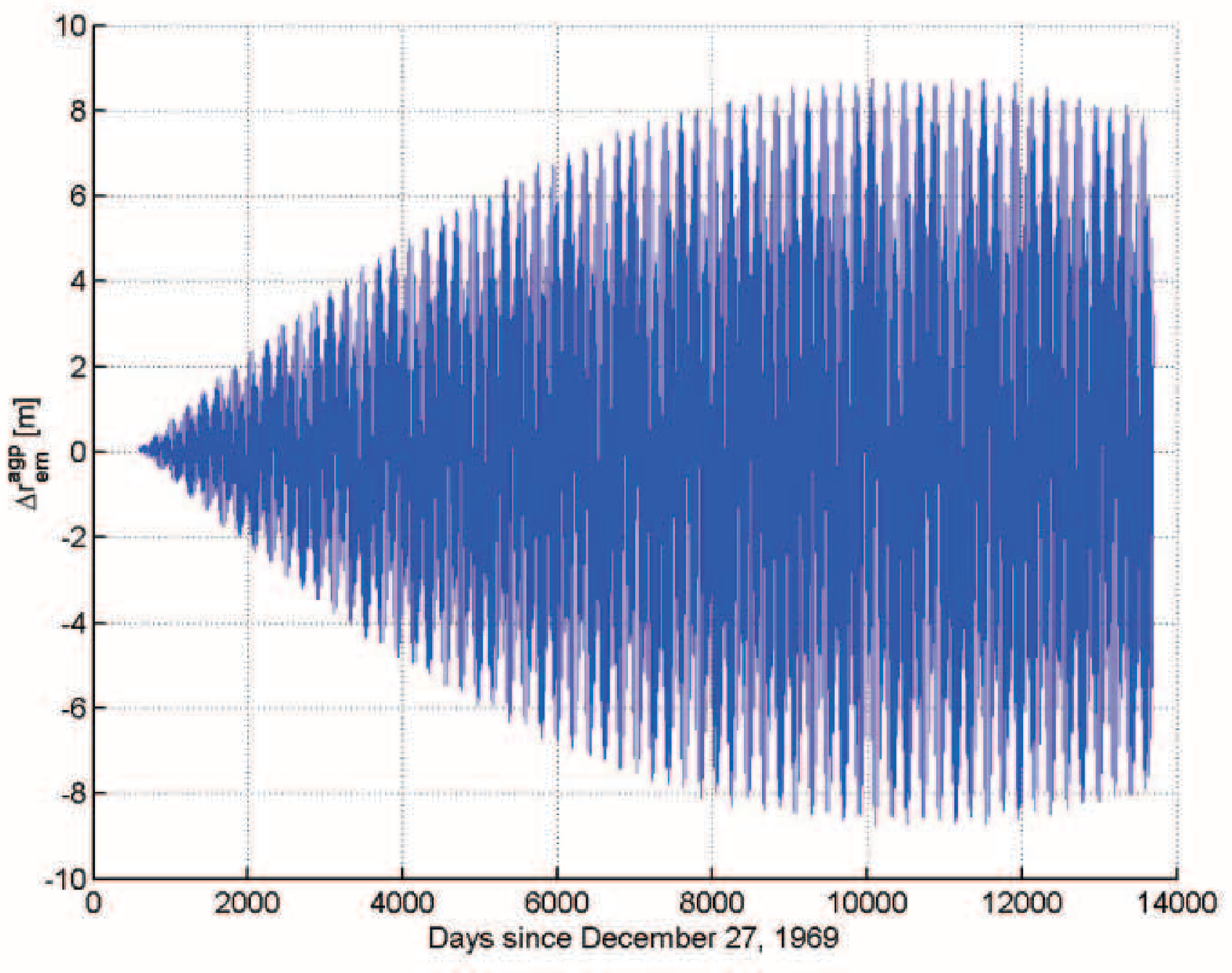}  
\end{minipage} 
\hfill
\begin{minipage}[b]{.46\linewidth}
\centering \psfig{width=7.4cm,     file=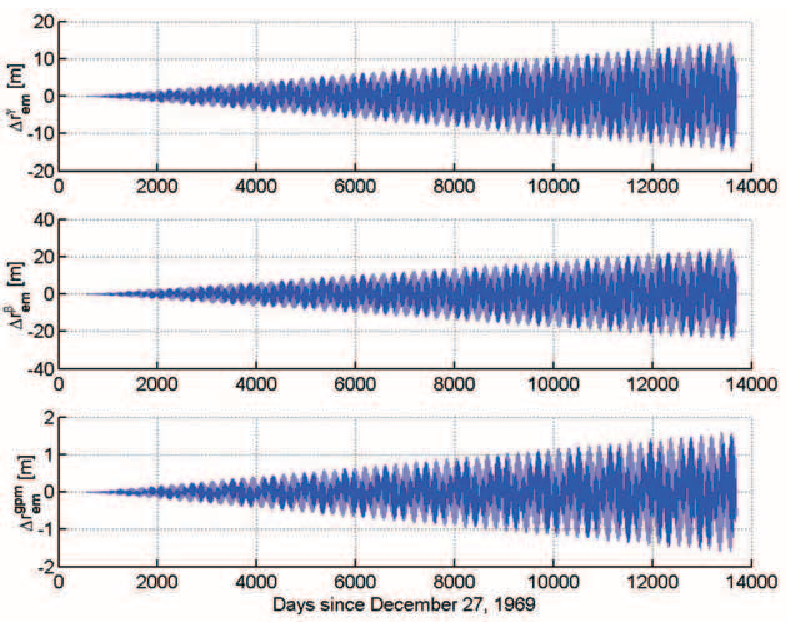}
\end{minipage}
\caption{(a) Sensitivity of LLR with respect to $\dot G / G$ assuming $\Delta \dot G / G = 8\cdot 10^{-13}\ {\rm yr}^{-1}$.
(b) Sensitivity of LLR with respect to space curvature $\gamma$, non-linearity couplings $\beta$ and geodetic precession using their present LLR accuracy (see Table 2) as perturbation value.
 \label{gpar_par1}}
\vskip -5pt 
\end{figure*}
%**********************************

%*********************************
\begin{figure*}[ht]
\begin{minipage}[b]{.46\linewidth}
\centering \psfig{width=7.4cm,     file=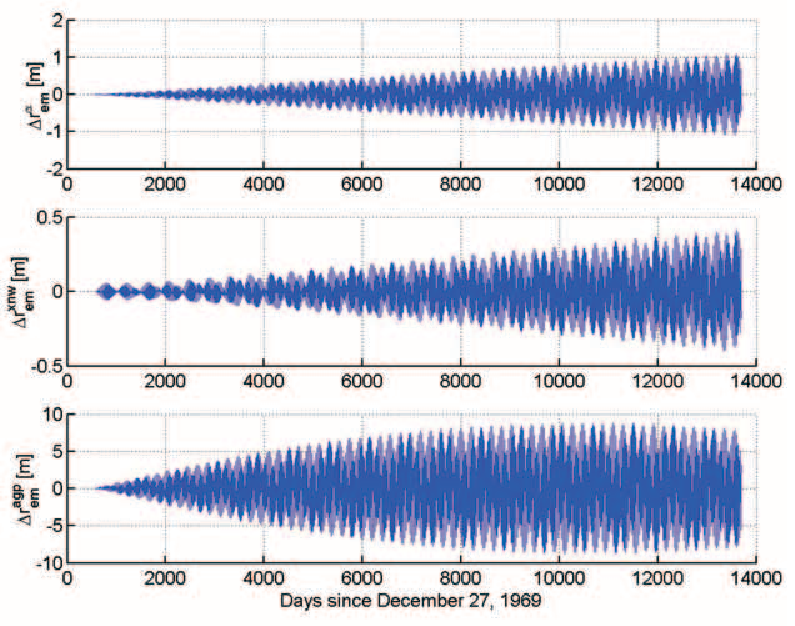}  
\end{minipage} 
\hfill
\begin{minipage}[b]{.46\linewidth}
\centering \psfig{width=7.4cm,     file=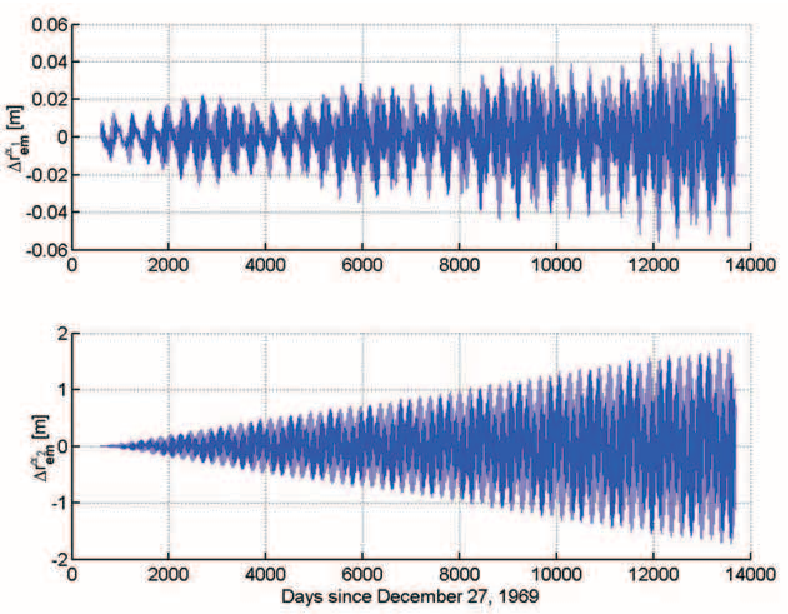}
\end{minipage}
\caption{(a) Sensitivity of LLR with respect to Yukawa interaction parameter $\alpha$, equivalence principle violation $\eta$ and time-variable gravitational constant $\dot G / G$ using their present LLR accuracy (see Table 2) as perturbation value.
(b) Sensitivity of LLR with respect to preferred-frame effects $\alpha_1$ and $\alpha_2$ using their present LLR accuracy (see Table 2) as perturbation value.
 \label{par2_par3}}
\vskip -5pt 
\end{figure*}
%**********************************

%*********************************
\begin{figure*}[ht]
\begin{minipage}[b]{.46\linewidth}
\centering \psfig{width=7.4cm,     file=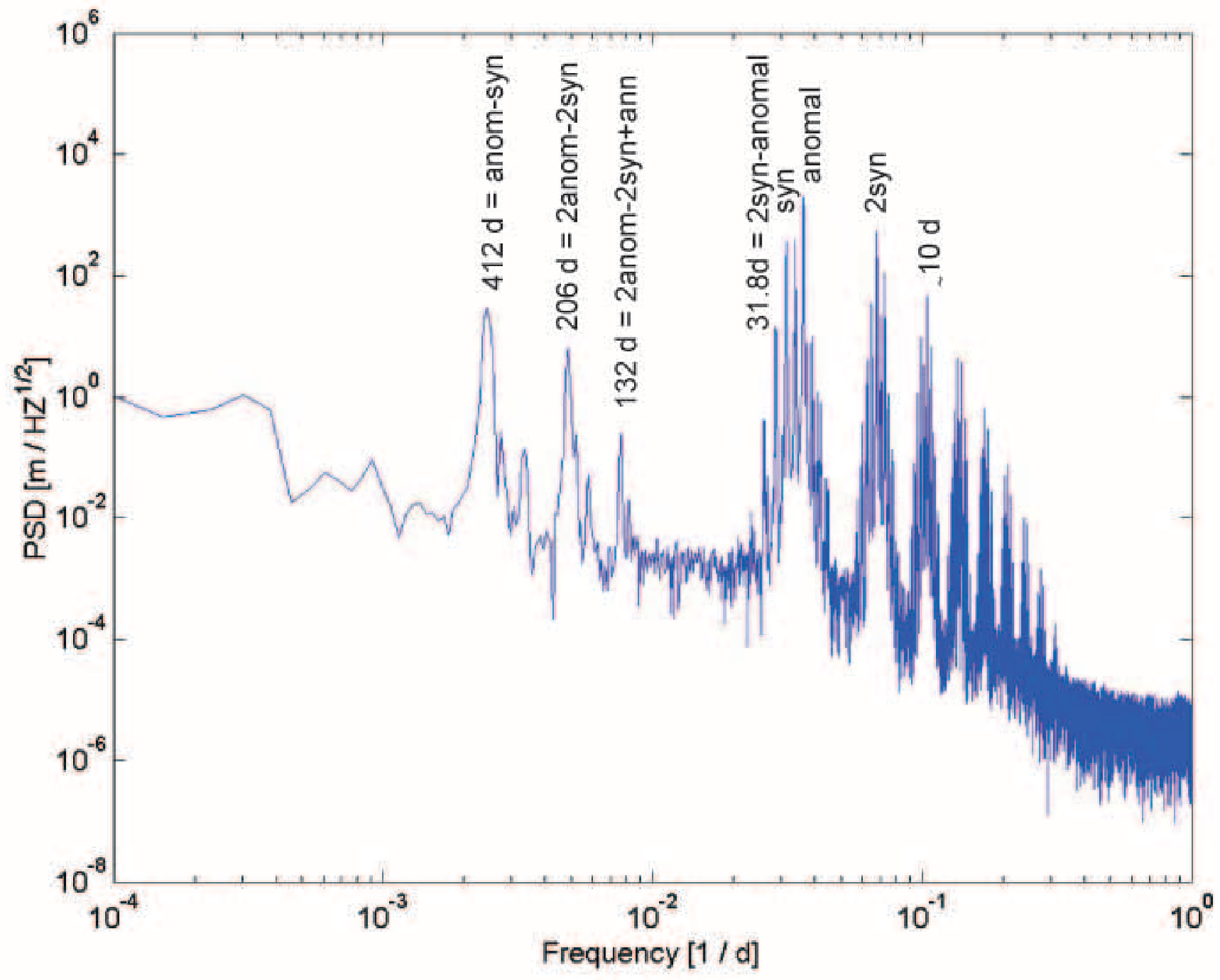}  
\end{minipage} 
\hfill
\begin{minipage}[b]{.46\linewidth}
\centering \psfig{width=7.4cm,     file=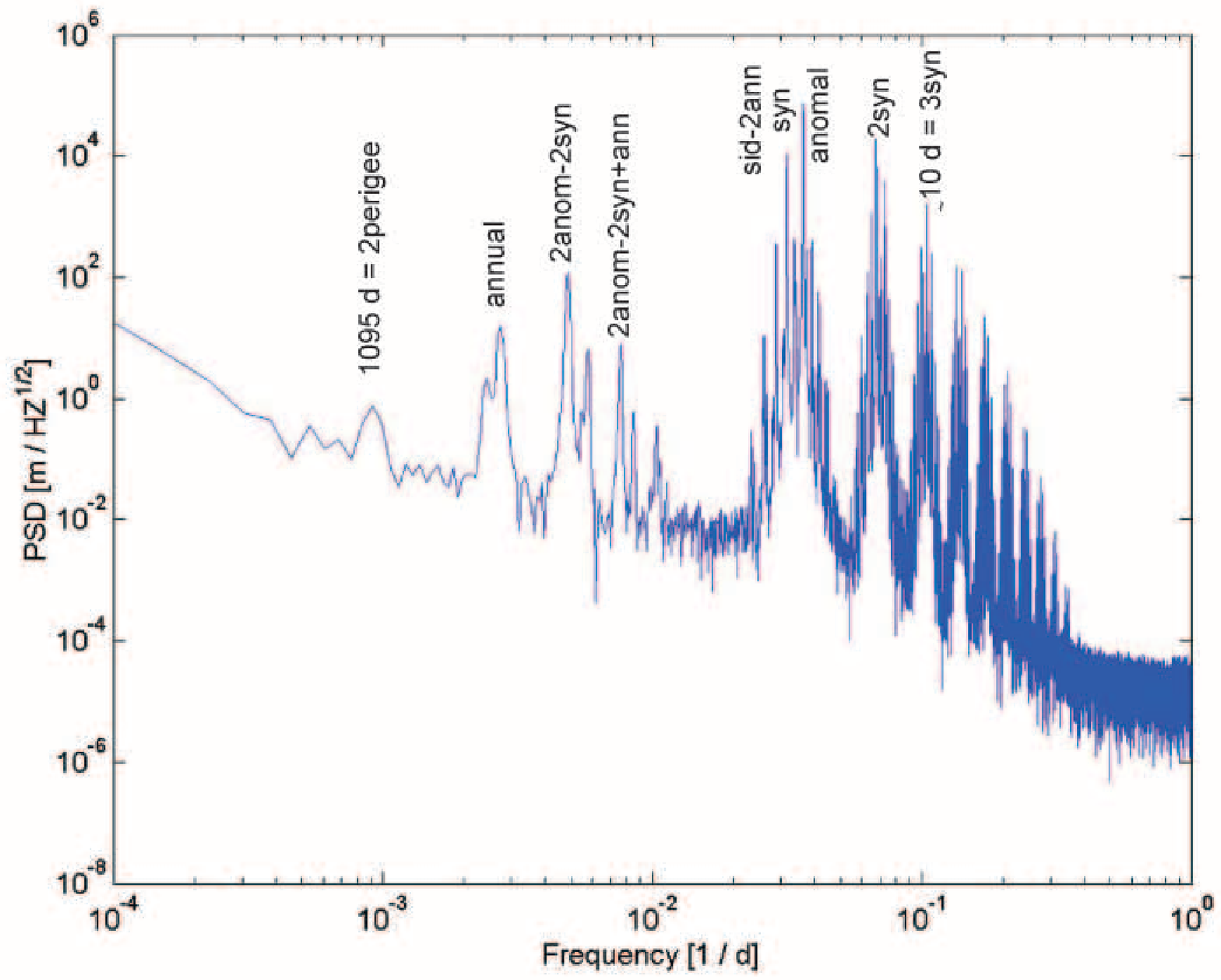}
\end{minipage}
\caption{
(a) Power spectrum of a possible equivalence principle violation  assuming $\Delta (m_G / m_I) \approx 10^{-13}$.
 (b) Power spectrum of the effect of $\dot G / G$ in the Earth-Moon distance assuming $\Delta \dot G / G 
  = 8\cdot 10^{-13}\ {\rm yr}^{-1}$.
 \label{gspec_espec}}
\vskip -5pt 
\end{figure*}
%**********************************

%*********************************
\begin{figure*}[ht]
\begin{minipage}[b]{.46\linewidth}
\centering \psfig{width=7.4cm,     file=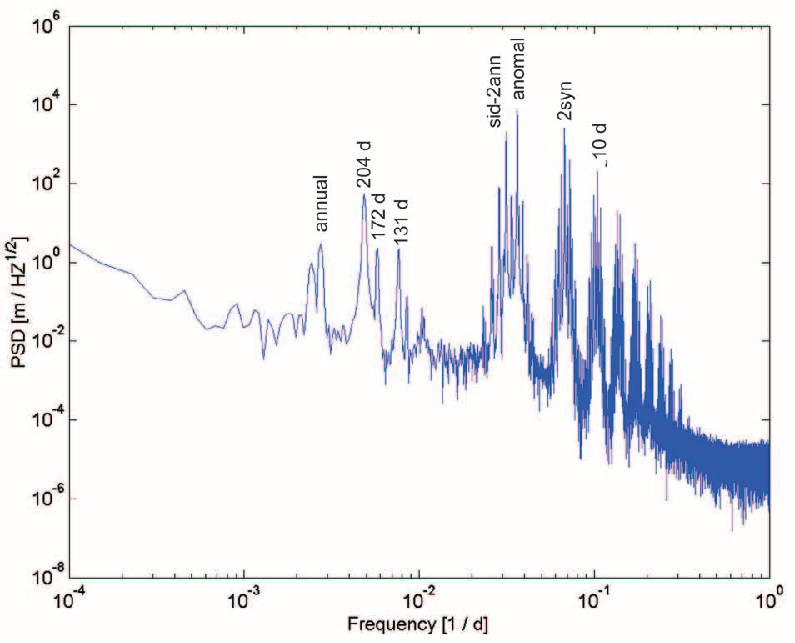}  
\end{minipage} 
\hfill
\begin{minipage}[b]{.46\linewidth}
\centering \psfig{width=7.4cm,     file=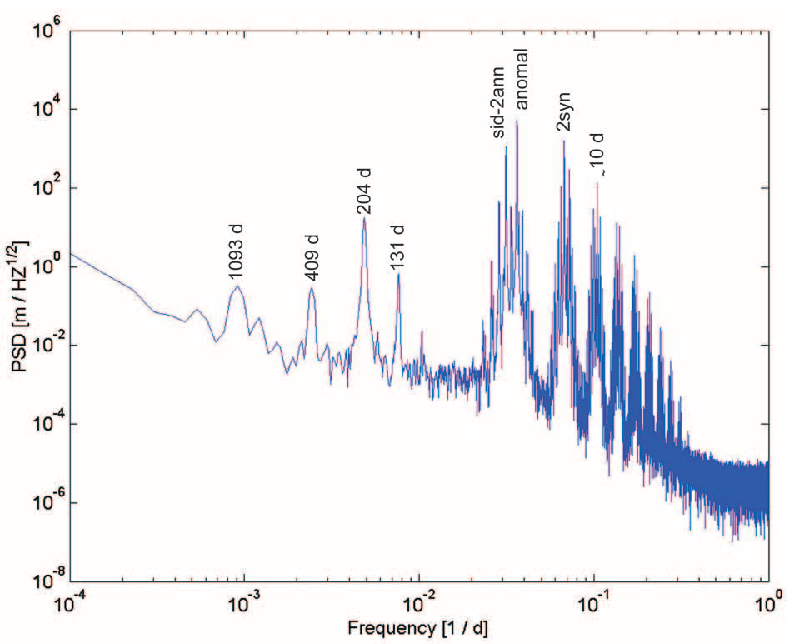}
\end{minipage}
\caption{(a) Power spectrum of an additional (deviating from Einstein's value) geodetic precession assuming 
  $\Delta gpm  = 10^{-2}$.
(b) Power spectrum of a possible Yukawa term using $\Delta \alpha 
  = 2\cdot 10^{-11}$.
 \label{gpm_yukspec}}
\vskip -5pt 
\end{figure*}
%**********************************
 
 %*********************************
\begin{figure*}[ht]
\begin{minipage}[b]{.46\linewidth}
\centering \psfig{width=7.4cm,     file=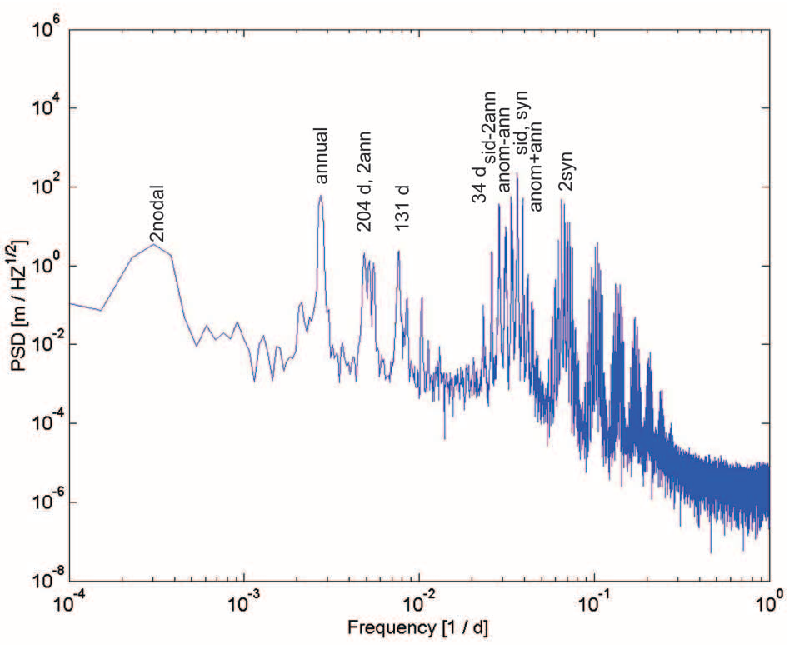}  
\end{minipage} 
\hfill
\begin{minipage}[b]{.46\linewidth}
\centering \psfig{width=7.4cm,     file=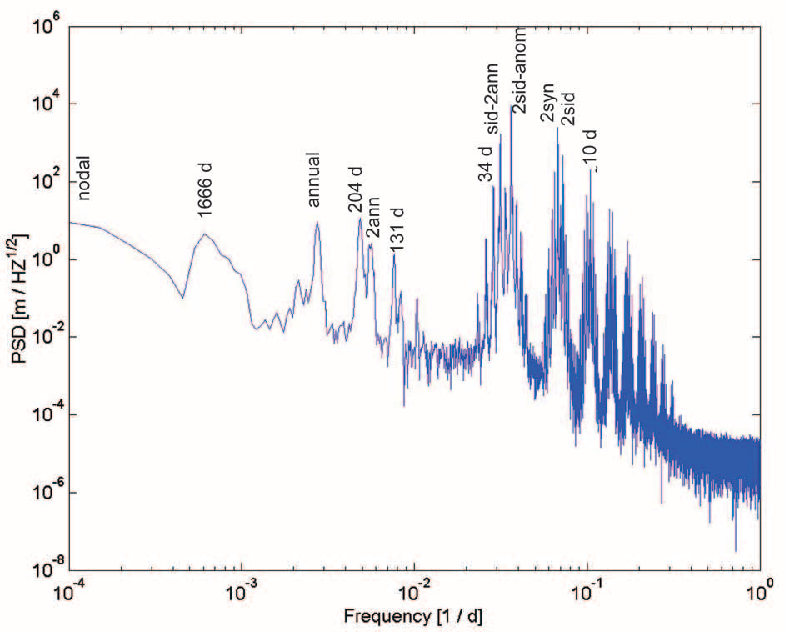}
\end{minipage}
\caption{(a) Power spectrum of a possible preferred-frame effect by $\alpha_1$ assuming $\Delta \alpha_1 
  = 9\cdot 10^{-5}$.
(b) Power spectrum of a possible preferred-frame effect by $\alpha_2$ assuming $\Delta \alpha_2 
  = 2.5\cdot 10^{-5}$.
 \label{a1_a2spec}}
\vskip -5pt 
\end{figure*}
%**********************************

\section{Sensitivity Study}
 
As indicated in Eqs.~(1)-(2), LLR is affected in various ways and at various levels by relativity. Relativity enters the equation of motion, i.e.\ the orbit and the rotation of the Moon.  More precisely, the
Earth-Moon system behaves according to relativity. But also the light propagation and the transformations between the reference and time frames which are used have to be modeled in agreement with general relativity. The lunar measurements contain the summed signal of all effects in one, so that the separation of the individual effects is a big challenge. To better understand what the individual contributions of the different relativistic effects are, sensitivity studies have been carried out, as 
\begin{equation}
\Delta r_{em}^p = \frac{\delta r_{em}}{\delta p}\Delta p.
\end{equation}
$\Delta r_{em}^p$ is the perturbation of the Earth-Moon distance caused by a  parameter $p$ which is one of the relativistic parameters described in Section 2. $\delta r_{em}/\delta p$ is the partial derivative of the Earth-Moon distance with respect to $p$; it is obtained by numerical differentiation. $\Delta p$ is a small value indicating the variation in $p$, here we have used the present realistic error as derived from LLR analyses (see Table~2).  As an example, Fig.\ \ref{gpar_par1}a represents the sensitivity of the Earth-Moon distance with respect to a possible temporal variation of the gravitational  constant in the order of $8\cdot 10^{-13}$ yr$^{-1}$, the present accuracy of that parameter. It seems as if perturbations of up to 9 meters are still caused, but this range (compared to the ranging accuracy at the cm level) can not fully be exploited, because the lunar tidal acceleration perturbation is similar. Fig.\ \ref{gpar_par1}b, Fig.\ \ref{par2_par3}a and Fig.\ \ref{par2_par3}b show the results of corresponding investigations for all relativistic parameters which were investigated during the present study (i.e.\ without the parameters discussed in items 7--9 of the previous Section). The perturbations vary between 5 cm ($\alpha_1$) and 25 m ($\beta$) which indicates that the former parameter is determined quite well from LLR data, as the sensitivity values are close to the observation accuracy, whereas the latter is only poorly determined because of its high correlation with the Newtonian orbit perturbations. Nevertheless, the continuation of LLR over a longer time span will help to further de-correlate the various parameters.
 
To better understand those couplings, corresponding power spectra have been computed. The largest periods for the EP-parameter are shown in Fig.\ \ref{gspec_espec}a and for $\dot G / G$  in  Fig.\ \ref{gspec_espec}b. Obviously many periods are affected simultaneously, because the perturbations, even if caused by a single beat period only (e.g.\ the synodic month for $\eta$ parameter\footnote{Note, here and throughout the paper, the relation $m_G / m_I = 1 + \eta (U/M c^2)$ has been used equivalently, where the second term describes the self energy of a body, cf.\ Williams et al.\ 2005b.}), change the whole lunar orbit (and rotation) and therefore excite further frequencies.  
For comparisons also the spectra of the geodetic precession $gpm$ and the Yukawa coupling parameter $\alpha$ are indicated (Fig.\ \ref{gpm_yukspec}a and Fig.\ \ref{gpm_yukspec}b). Again a different combination of periods is visible. As before mainly the monthly (e.g.\ sidereal, synodic, anomalistic), half-monthly, etc., periods dominate, but longer periods (low frequencies) are also present, e.g.\ annual, 3 years or combinations of the monthly and annual frequencies.
Similar pictures are obtained when considering the preferred-frame parameters  $\alpha_1$ and $\alpha_2$ (Fig.\ \ref{a1_a2spec}a and Fig.\ \ref{a1_a2spec}b). A huge variety  of periods show up again, but they differ partly from each other (note, e.g., the very low-frequency contributions).   The spectra of $\gamma$ and $\beta$ (not shown here) are very similar to the geodetic precession spectrum. Although there are big similarities between the various spectra, the typical properties of each can be used to identify and separate the different effects and to determine corresponding parameters.

\section{Results}

The global adjustment of the model by least-squares-fit procedures gives improved values for the estimated parameters and their formal standard errors, while consideration of parameter correlations obtained from the covariance analysis and of model limitations lead to more 'realistic' errors. Incompletely modeled solid Earth tides, ocean loading or geocenter motion, and uncertainties in values of fixed model parameters have to be considered in those estimations.  See Table 2 for the determined values for the relativistic quantities and their realistic errors.   

The EP-parameter $\eta$ $\left( =(6 \pm 7) \cdot  10^{-4}\right)$ benefits most from highest accuracy over a sufficient long time span (e.g.\ one year) and a  good data coverage over the synodic month, as far as possible. Also any observations reducing the asymmetry  about the quarter Moon phases (compare Fig.\ \ref{ann_syn}b) would improve the fit of $\eta$.  The improvement of the EP parameter was not so big since 1999, as the LLR RMS residuals increased a little bit in the past years, see Fig.\ \ref{sid_rms}b. The reason for that increase is not completely understood and has to be investigated further.

In combination with the recent value of the space-curvature parameter $\gamma_{\rm Cassini}$ $\left(\gamma -1 = (2.1 \pm 2.3) \cdot 10^{-5}\right)$ derived from Doppler measurements to the Cassini spacecraft  
(Bertotti et al.\ 2003), the non-linearity parameter $\beta$ can be determined by applying the relationship $\eta=4\beta-3-\gamma_{\rm Cassini}$. One obtains $\beta - 1 = (1.5 \pm 1.8) \cdot 10^{-4}$ (note that using the EP test to determine parameters $\eta$ and $\beta$ assumes that there is no composition-induced EP violation).
The LLR result for the space-curvature parameter $\gamma$ as determined from the EIH equations is less accurate than the results derived from other measurements, because its effect is very similar to the Newtonian orbit perturbations.

For the temporal variation of the gravitational constant, $\dot G / G =  (6\pm 8)\cdot 10^{-13}$ has been  obtained, where the formal standard deviation has been scaled by a factor 3  to yield the given value. This  parameter benefits most from the long time span of LLR data and has experienced  the biggest improvement over  the past years (cf.\ M\"uller et al.\ 1999). 

For the estimation of the de Sitter precession of the lunar orbit, a Coriolis-like term is added to the equation of motions, which adds the precession effect as predicted by Einstein for a second time. This term is scaled by a parameter called $gpm$ which has to give 0 if Einstein is correct. $gpm = 0$ is obtained with an accuracy of
about 1 percent.

The Yukawa coupling parameter $\alpha$ has been determined by adding corresponding perturbation terms to the equations of motion, where the partials were computed by numerical differentiation. The recently determined value shows small deviations from the expected value; it will be further investigated in future.

The preferred-frame parameters $\alpha_1$ and $\alpha_2$ can either be determined by extending the equations of motion  or by adding analytical terms to the Earth-Moon distance. In both cases quite similar results are obtained (see M\"uller et al.\ 1996, 1999). Recent determinations are given in Table 2. 

The Mansouri-Sexl parameters $\zeta_0$ and $\zeta_1$ as well as the quantity indicating a possible EP-violating coupling with dark matter were not investigated during our present studies; the values given in Table 2 are taken from M\"uller et al.\ (1999).

A further question to be considered in more detail in future is the right combination of the parameters. That means, shall all relativistic parameters be estimated together in one global adjustment, or each one alone (together with the parameters of the standard solution)? We carried out several tests considering the correlation of the relativistic quantities with each other, but also with the 'classical' ones, e.g.\ with the orbital parameters or site velocities (Koch 2005). It is too early to make a final decision. On the one hand 'over-estimation' of an effect has to be avoided, on the other hand  over-constraining by fixing too many parameters should also be avoided. 

Final results for all relativistic parameters obtained from the IfE analysis are shown in Table 2 (see also M\"uller et al. 2005). The realistic errors are comparable with those obtained in other recent investigations, e.g.\ at JPL (see  Williams et al.\ 1996, 2004a, 2004b, 2005b).

\begin{table}[htb]
\caption{Determined values for the relativistic quantities and their realistic errors.}
\begin{center}
\begin{tabular}{|c|c|}
\hline
 Parameter    &  Results \\ \hline\hline
  Equivalence\ Principle parameter\ $\eta$  & $(6 \pm 7) \cdot 10^{-4} $ \\ \hline
 Metric parameter \ $\gamma - 1$ & $(4 \pm 5) \cdot 10^{-3} $ \\ \hline
 Metric parameter \ $\beta - 1$: direct measurement  & $(-2\pm 4) \cdot 10^{-3} $ \\ 
and from $\eta=4\beta-3-\gamma_{\rm Cassini}$ & $(1.5\pm 1.8) \cdot 10^{-4} $ \\ \hline
Time varying\ gravitational \ constant\ $\dot G / G$ [${\rm yr}^{-1}$] & $(6\pm 8)\cdot 10^{-13} $  \\ \hline
 Differential\ geodetic\ precession\ $\Omega_{\rm GP}$ - $\Omega_{\rm deSit}$ [''/cy] & $(6\pm 10)\cdot 10^{-3} $   \\ \hline
  Yukawa\ coupling\ constant\ $\alpha$ (for $\lambda = 4\cdot 10^5\, {\rm km})$
 & $(3\pm 2) \cdot 10^{-11} $ \\ \hline
  `Preferred frame' parameter $\alpha_1$  & $(-7\pm 9) \cdot 10^{-5} $
 \\ \hline
 `Preferred frame' parameter $\alpha_2$  & $(1.8 \pm 2.5) \cdot 10^{-5} $
 \\ \hline
 Special\ relativistic\ parameters $\zeta_1 - \zeta_0 - 1$  & $(-5\pm 12) \cdot 10^{-5} $
 \\ \hline
  Influence\ of dark matter $\delta g_{\rm galactic}$ [cm/s$^2$] & $(4\pm 4) \cdot 10^{-14} $
 \\ \hline
\end{tabular}
\end{center}
\vskip -10pt
\end{table}

\section{Further Applications}

\begin{figure}[tbh]
  \begin{center}
  \epsfig{width=7.4cm,     file=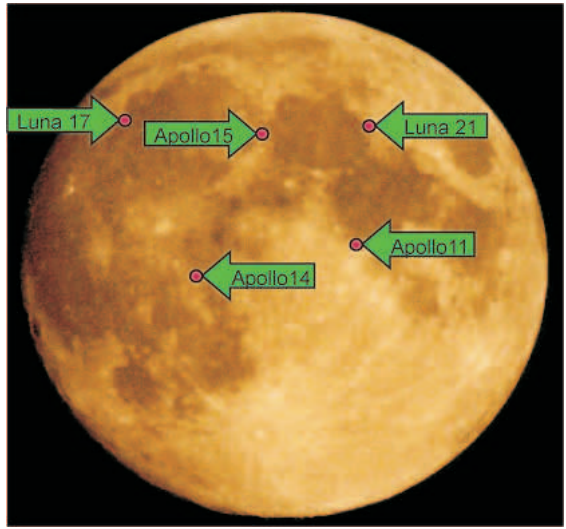}
  \end{center}\vskip -10pt
  \caption{Distribution of retro-reflectors on the lunar surface.} 
\label{refmoon}
\vskip -10pt
 \end{figure}

In addition to the relativistic phenomena discussed above, more effects related to lunar physics, geosciences, and geodesy  can be investigated. The following items are of special interest (see also M\"uller et al. 2005), as they touch recent activities in the afore-mentioned disciplines:
\begin{enumerate}

\item {\it Celestial reference frame:} A dynamical realization of the International Celestial Reference System (ICRS) by the lunar orbit is obtained ($\sigma$~=~0.001") from LLR data. This can  be compared and analyzed with respect to the kinematical ICRS from Very Long Baseline Interferometry (VLBI). Here, the very 
good long-term stability of the lunar orbit is of great advantage.  

\item {\it Terrestrial reference frame:} The results for the station coordinates and velocities, which are estimated simultaneously in the standard solution, contribute to the realization of the international terrestrial reference frame, e.g.\ to the last one, the ITRF2000.

\item {\it Earth rotation:} LLR contributes, among others, to the determination of long-term nutation parameters, where again the stable, highly accurate orbit and the lack of non-conservative forces from air-drag or solar radiation pressure (which affect satellite orbits substantially) is very convenient. Additionally UT0 and VOL values are computed, which stabilize the combined EOP series, especially in the 1970s when less data from other space geodetic techniques were available. The precession rate is another example in this respect.

\item {\it Relativity:} As discussed in the previous sections, the dedicated investigation of Einstein's theory of relativity is of major interest.  With an improved accuracy of the LLR measurements and the modeling (see next Section) the investigation of further effects (e.g.\ the Lense-Thirring precession) or those of alternative theories might become possible.

\item {\it Lunar physics:} By the determination of the libration angles of the Moon, LLR gives access to underlying processes affecting lunar rotation (e.g.\ Moon's core, dissipation), cf. Williams et al.\ (2005a). A better distribution of the retro-reflectors on the Moon (see  Fig.\ \ref{refmoon}) would be very helpful.

\item {\it Selenocentric reference frame:} The determination of a selenocentric reference frame, the combination with high-resolution images and the establishment of a better geodetic network on the Moon is a further big item, which then allows accurate lunar mapping.

\item {\it Earth-Moon dynamics:} The mass of the Earth-Moon system, the lunar tidal acceleration, possible geocenter variations and related processes as well as further effects can be investigated in detail. 
 
\item {\it Time scales:} The lunar orbit can also be considered as a long-term stable clock so that LLR can be used for the independent realization of time scales, which can then be compared or combined with other determinations. 
\end{enumerate}

Those features shall be addressed in the future, when more and better LLR data are available and the analysis models have been improved to the mm level, see next Section.

\section{Model and Observation Refinement}

To exploit the full available potential of LLR, the theoretical models as well as the measurements require optimization. Using the 3.5 m telescope at the APOLLO site in New Mexico, USA, a mm-level ranging becomes possible. To allow an order of magnitude gain in determination of various quantities in the complete LLR solution, the current models have to be up-dated according to the IERS conventions 2003, and  made compatible with the IAU 2000 resolutions. 
This requires, e.g., to better model
\begin{itemize}
\item higher degrees of the gravity fields of Earth and Moon and their couplings;
\item the effect of the asteroids (up to 1000);
\item relativistically consistent torques in the rotational equations of the Moon; 
\item relativistic spin-orbit couplings;
\item torques caused by other planets like Jupiter;
\item the lunar tidal acceleration with more periods (diurnal and semi-diurnal); 
\item ocean and atmospheric loading by updating the corresponding subroutines;
\item nutation using the recommended IAU model; 
\item the tidal deformation of Earth and Moon;
\item Moon's interior (e.g.\ solid inner core) and its coupling to the Earth-Moon dynamics.
\end{itemize}
Besides modeling, the overall LLR processing shall be optimized. The best strategy for the data fitting procedure needs to be explored for (highly) correlated parameters.

Finally LLR should be prepared for a renaissance of lunar missions where transponders or new retro-reflectors may be deployed on the surface of the Moon which would enable many pure SLR stations to observe the Moon. NASA is planning to return to the Moon by 2008 with Lunar Reconnaissance Orbiter (LRO), and later with robotic landers, and then with astronauts until 2018.  The primary focus of these planned missions will be lunar exploration and preparation for trips to Mars, but they will also provide opportunities for science, particularly if new reflectors are placed at more widely separated locations than
the present configuration (see Fig.\ \ref{refmoon}). New installations on the Moon would give stronger determinations of lunar rotation and tides. New reflectors on the Moon would provide additional accurate surface positions for cartographic control (Williams et al.\ 2005b), would also aid navigation of surface vehicles or spacecraft at the Moon, and they also would contribute significantly to research in fundamental and gravitational physics, LLR-derived ephemeris and lunar rotation. Moreover in the case of co-location of microwave transponders, the connection to the VLBI system may become possible which will open a wide range of further activities such as frame ties.

\section{Conclusions}

LLR has become a technique for measuring a variety of relativistic gravity parameters with unsurpassed precision. Sensitivity studies have been performed to estimate the order of magnitude of relativistic effects on lunar ranges and the potential capability to better determine certain relativistic features. Spectral analyses showed the typical frequencies related to each effect, indicating as well, how (highly) correlated parameters might be separated. No violations of general relativity are found in our investigations.  Both the weak and strong forms of the EP are verified, while strong empirical limitations on any inverse square law violation, time variation of $G$, and preferred frame effects are also obtained. 

LLR continues as an active program, and it can remain as one of the most important tools for testing Einstein's general relativity theory of gravitation if appropriate observations strategies are adopted and if the basic LLR model is further extended and improved down to the millimeter level of accuracy. The deployment of transponders on the Moon would considerably improve the performance for lunar ranging applications. Lunar science, fundamental physics, control networks for surface mapping, and navigation would benefit. Demonstration of active devices would prepare the way for very accurate ranging to Mars and other solar system bodies.

\vspace{0.3cm}\noindent
{\bf\ Acknowledgments.}
Current LLR data is collected, archived and distributed under the auspices of the International Laser Ranging Service (ILRS).  All former and current LLR data is electronically accessible through the European Data Center (EDC) in Munich, Germany and the Crustal Dynamics Data Information Service (CDDIS) in Greenbelt, Maryland.  The following web-site can be queried for further information: {\tt http://ilrs.gsfc.nasa.gov}.  We also acknowledge with thanks, that the more than 35 years of LLR data, used in these analyses, have been obtained under the efforts of personnel at the Observatoire de la C\^ote d'Azur, in France, the LURE Observatory in Maui, Hawaii, and the McDonald Observatory in Texas.

A portion of the research described in this paper was carried out at the Jet Propulsion Laboratory of the California Institute of Technology, under a contract with the National Aeronautics and Space Administration.

\end{document}